\documentclass[aps,prl,twocolumn,footinbib,superscriptaddress,amsmath,amssymb,floatfix]{revtex4-2}
\usepackage{amsfonts,mathtools,dsfont}
\usepackage[T1]{fontenc}
\usepackage[utf8]{inputenc}
\usepackage{textgreek}
\usepackage{braket}
\usepackage[hypertexnames=false]{hyperref}
\usepackage{graphicx}
\usepackage{siunitx}
\usepackage[english]{babel}
\usepackage{xfrac}
\usepackage[version=4]{mhchem}

\usepackage{prv2al20_defs}

\newcommand\titlelowercase[1]{\texorpdfstring{\lowercase{#1}}{#1}}

\usepackage{xcolor}

\makeatletter
\def\maketitle{
\@author@finish
\title@column\titleblock@produce
\suppressfloats[t]}
\makeatother

\begin{document}
\newcommand{\thetitle}{Strange-metal behavior without fine-tuning in \texorpdfstring{\ce{PrV2Al20}}{PrV2Al20}}
\title{\thetitle}

\author{Marvin Lenk}
\email[E-mail: ]{mlenk@th.physik.uni-bonn.de}
\affiliation{Physikalisches Institut, Universit\"at Bonn, Nussallee 12, 53115 Bonn, Germany}

\author{Fei Gao}
\affiliation{Department of Physics and Astronomy, Rice University, Houston, Texas 77005, USA}

\author{Johann Kroha}
\email[E-mail: ]{kroha@th.physik.uni-bonn.de}
\affiliation{Physikalisches Institut, Universit\"at Bonn, Nussallee 12, 53115 Bonn, Germany}
\affiliation{\hbox{School of Physics and Astronomy, University of St. Andrews, North Haugh, St. Andrews, KY16 9SS, United Kingdom}}

\author{Andriy H. Nevidomskyy}
\email[E-mail: ]{nevidomskyy@rice.edu}
\affiliation{Department of Physics and Astronomy, Rice University, Houston, Texas 77005, USA}

\begin{abstract}
    Strange-metal behavior observed in the praseodymium-based heavy-fermion material \ce{PrV2Al20} has been tentatively interpreted in the framework of proximity to a quantum critical point (QCP) associated with quadrupolar ordering. Here, we demonstrate that an alternative, natural explanation exists without invoking a QCP, in terms of the unconventional nature of the quadrupolar Kondo effect taking place in non-Kramers ions. Using a combination of \textit{ab initio} density-functional theory calculations and analytical arguments, we construct a periodic Anderson model with realistic parameters to describe \ce{PrV2Al20}. We solve the model using dynamical mean-field theory preserving the model symmetries and demonstrate the non-Fermi liquid strange-metal behavior stemming from the two-channel nature of the quadrupolar Kondo effect. Our calculations provide an explanation for the puzzling temperature dependence in the magnetic susceptibility and provide a basis for analyzing future photoemission experiments.
\end{abstract}
\maketitle

In the past several decades, so-called strange metals and associated phenomena have garnered much attention. As the name suggests, these metals defy the standard Landau Fermi liquid paradigm~\cite{landau_theory_1956} in that they lack well-defined quasiparticles, as demonstrated by the unusual, non-Fermi liquid (NFL) power-law exponents in the electrical resistivity, specific heat, magnetic susceptibility, and other thermodynamic and transport quantities. A large number of strange metals have been found in the heavy-fermion materials when magnetic ordering is suppressed to zero temperature by pressure, doping, or magnetic field, resulting in a quantum critical point (QCP)~\cite{CeCu6-lohneysen1996,CeCoIn5-paglione2003,YRS-custers2003,NFL-review-gegenwart2008,Ybal-matsumoto2011}. In some instances, strong quantum fluctuations prevent the textbook Nozi\`eres' Fermi liquid state from forming, resulting in Kondo breakdown~\cite{Si2001,Coleman2001,senthil2004}. 

In the vast majority of heavy-fermion metals, the conduction electrons act to screen a spin-$\frac{1}{2}$ Kramers ion, 
such as Ce$^{3+}$~\cite{CeCu6-lohneysen1996,CeCoIn5-paglione2003} or Yb$^{3+}$~\cite{YRS-custers2003,Ybal-matsumoto2011}, resulting in a competition between the Kondo effect and incipient magnetic ordering of the localized moments~\cite{Doniach, Kroha_2017}. The situation is less clear in materials with non-Kramers ions, like praseodymium (Pr)-based PrPb$_3$ and PrPtBi. There, the 4f$^2$ electrons are well localized but only very weakly coupled to conduction electrons, making the Kondo scale $T_K$ too small to be observed.

The \textit{status quo} has changed dramatically with the discovery of
Pr$T_2$Al$_{20}$ ($T$=Ti, V) about a decade ago, where the large values of the Seebeck coefficient~\cite{Kuwai_thermopower_2013,Machida_thermopower_2015} and hyperfine coupling~\cite{Tokunaga2013}, much enhanced effective masses ($m^*/m_e \sim 20$ in PrTi$_2$Al$_{20}$ and $\sim 120$ in PrV$_2$Al$_{20}$~\cite{Shimura2015}), and direct evidence for the Abrikosov-Suhl resonance in photoemission~\cite{Matsunami2011} all indicate that Kondo hybridization plays an important role in these compounds. 
Intriguingly, non-Fermi liquid behavior is observed in PrV$_2$Al$_{20}$~\cite{sakai2011,Tsujimoto2015}, prompting the suggestions that this material may sit close to a putative QCP.
Although a precedent for a stoichiometric material perched in close proximity to a QCP exists, notably $\beta$-YbAlB$_4$~\cite{Ybal-matsumoto2011}, this is certainly unusual. 
A possible alternative to explaining the strange metal behavior in PrV$_2$Al$_{20}$ is based on the notion that the Kondo effect is unconventional in that no dipole moment is present in the $|\Gamma_3\ra$ crystal-electric-field (CEF) ground state of the Pr$^{2+}$ ion. The conduction electrons instead screen the quadrupolar moment of the non-Kramers doublet so that the conduction-electron spin represents two conserved screening channels protected by Kramers degeneracy.
The resulting two-channel Kondo (2CK) effect is known to harbor non-Fermi liquid physics~\cite{affleck1993} and this option has previously been investigated generally for quadrupolar Kondo materials and specifically in the context of \ce{PrV2Al20}~\cite{tsuruta_non-fermi_2015, inui_channel-selective_2020, patri_emergent_2020, patri_critical_2020, schultz_rise_2021}. However,
measurements of the magnetic susceptibility $\chi(T)$~\cite{sakai2011, araki2014} are at odds with the expected behavior for a standard 2CK effect~\cite{affleck1993}. Thus, the question remains open whether the 2CK effect alone can explain the strange-metal behavior or if it is necessary to involve competing quadrupolar order and the QCP scenario.

In this Letter, we use  \emph{ab initio} band-structure calculations combined with analytical insight to formulate a two-channel Anderson lattice model that captures the salient features of 1-2-20 compounds, focusing in particular on \ce{PrV2Al20}. We then solve the model using dynamical mean-field theory (DMFT), finding that the spectral density shows a clear Abrikosov-Suhl resonance while the quasiparticle decay rate exhibits NFL scaling that approaches 2CK behavior. Surprisingly, we find that the magnetic susceptibility is dominated by the first excited CEF multiplet (the magnetic $\Gamma_5$ triplet) since it couples linearly to the magnetic field, in contrast to the quadratic coupling of the quadrupolar $\Gamma_3$ ground-state doublet. The Kondo physics of the triplet contribution is fundamentally different in that it gives rise to an intermediate impurity spin-$1$ Kondo effect, where the
conduction-electron spin-$\frac{1}{2}$ is actively participating in the screening instead of acting as a channel.
This crossover~\cite{Note1} between an impurity spin-1 Kondo effect at intermediate temperatures and 2CK behavior at low temperatures is responsible for the experimentally observed strange-metal behavior in \ce{PrV2Al20}~\cite{sakai2011} without involving the notion of it being fine-tuned to a QCP.

\begin{figure}[tb]
    \centering
    \includegraphics[width=\columnwidth]{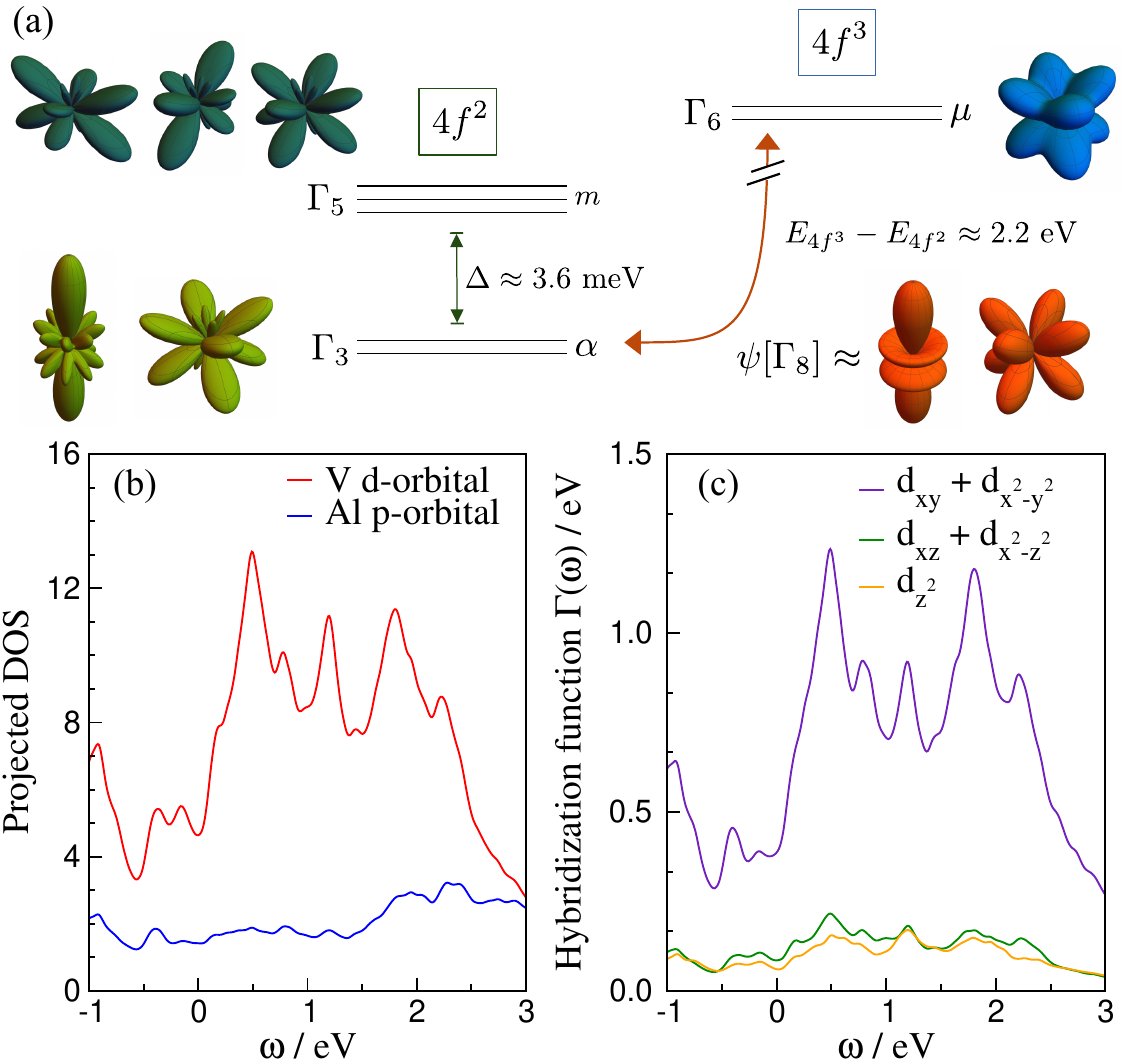}
    \caption{
    a) CEF-Scheme of the states involved in the valence fluctuations of the \ce{Pr} atoms. b) Projected density of states (DOS) of \ce{V} $d$ orbitals and \ce{Al} $p$ orbitals normalized to the number of \ce{Pr} neighbors instead of the number of atoms in a unit cell. c) Contribution to the hybridization function from different \ce{V} $d$ orbitals (c.f.~Eq.~\eqref{eq:hybfunc}). 
    \label{fig:CF_scheme_hybf_dxy}}
\end{figure}

\textit{Valence fluctuations and CEF excitations.} --
The experimentally determined CEF ground state of the Pr$^{2+}$ ion is the $|\Gamma_3\pm\ra$ doublet of the cubic $T_d$ point group (see Supplementary Materials~\cite{supplement} and Refs.~\cite{onimaru2016, momma_vesta_2011, schiller_mixed-valent_1998, kusunose_interplay_2005} cited therein for details). The valence fluctuations are between this $4f^2$ ground state and the excited $4f^3$ and $4f^1$ Kramers states. In order to determine the dominant channel, we have performed \textit{ab initio} density functional theory (DFT) calculations using the full-potential augmented plane-wave method implemented in the Wien2k package~\cite{wien2k}, with a fixed number of $f$-electrons in the core (the `open core' method, see \cite{supplement} for details). We found the energy difference to the $|4f^3\ra$ state, $\Delta E_{2 \leftrightarrow 3} = E_{4f^3} - E_{4f^2} \approx \SI{2.2}{\eV}$, to be significantly lower than the energy of the excited $|4f^1\ra$ level ($\Delta E_{2 \leftrightarrow 1} \approx \SI{3.8}{\eV}$). We, therefore, adopt $4f^2 \leftrightarrow 4f^3$ as the dominant Kondo hybridization channel -- to the best of our knowledge, this is the first such calculation of its kind. 
According to Cox and Zawadowski~\cite{cox1998}, $\Gamma_6$ is the ground-state Kramers doublet (due to spin degeneracy) of the $4f^3$ multiplet, which we adopt for our calculation~\cite{Note2}. 

The aforementioned transitions are mediated via conduction electrons projected onto $\Gamma_8$ states at the \ce{Pr} atom positions, since the product of the irreducible representations (irreps) $\Gamma_6 \otimes \Gamma_3 = \Gamma_8$. Note that  the $\Gamma_8$ state is 4-fold degenerate, labeled by the isospin $\ipsspin=\pm$ and the Kramers spin $\ichan=\up,\,\dn$.  
The valence fluctuations and the corresponding multiplets are shown schematically in Fig.~\ref{fig:CF_scheme_hybf_dxy}(a). As we demonstrate below, the first excited $4f^2$ CEF state ($\Gamma_5$ triplet) plays a crucial role in the magnetic susceptibility. Experimental data suggest a CEF splitting of $\Delta / k_B \approx \SI{40}{\kelvin}$~\cite{sakai2011}.

\textit{Two-channel nature of the Kondo effect.}
The non-Kramers $\Gamma_3$ doublet has no magnetic dipole but a non-vanishing quadrupole~\cite{cox1998,hattori2014} and octupole moment~\cite{hattori2014,hattori2016}.
The conduction electrons screen this state in an unconventional manner, dubbed the quadrupolar Kondo effect, which was first studied by D. Cox and A. Zawadowski~\cite{cox1987,cox1998} in the context of uranium-based heavy-fermion materials.
They made the crucial observation that the Kramers degeneracy inherent in the spin-$1/2$ conduction electrons provides two degenerate channels for Kondo screening, thus resulting in a two-channel Kondo (2CK) effect which is known to exhibit non-Fermi liquid behavior below the Kondo temperature $T_K$~\cite{AndreiDestri_1984,EmeryKivelson_1992}. 
In this Letter, we build a microscopically faithful model of this effect in \ce{PrV2Al20}, introduced below, and use DMFT to solve it numerically.
\begin{figure*}[t!]
    \centering
    \includegraphics[width=\textwidth]{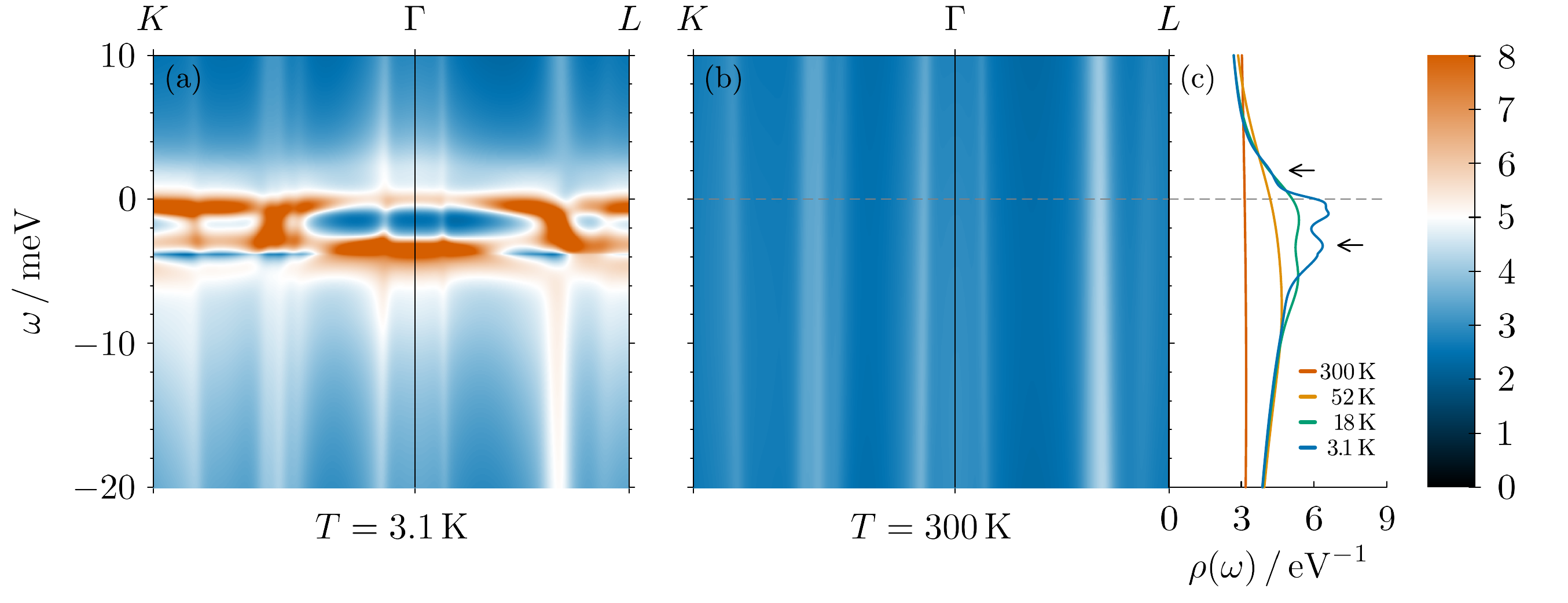}
        \caption{
        (a), (b): Momentum-dependent, renormalized spectral functions as obtained from DMFT, traced over the V $d_{xy + x^2-y^2}$ and Pr $4f$ bands, for temperatures $T=\SI{3.1}{\kelvin}$ and $T=\SI{300}{\kelvin}$. 
        (c) Momentum-integrated density of states for different temperatures. Arrows indicate features at $\omega = -\SI{3.2}{\milli\eV}$ and $\omega = \SI{2}{\milli\eV}$. 
        \label{fig:spectrum}}
\end{figure*}

\textit{Model.} -- 
We construct a periodic Anderson model (PAM) to capture the $4f^2 \leftrightarrow 4f^3$ valence fluctuations.
It is most naturally written in terms of Hubbard operators~\cite{hubbard_electron_1964}, which we conveniently express by auxiliary particle operators~\cite{coleman_new_1984, barnes_new_1977, abrikosov_electron_1965, kroha_NCA_1996},
\begin{align}
    a^\dagger_{j,\ichan} \ket{0} &= \ket{j;4f^3 \; \Gamma_6, \ichan},\label{eq:slaveboson}\\
    f^\dagger_{j,\ipsspin} \ket{0} &= \ket{j;4f^2 \; \Gamma_3, \ipsspin},\label{eq:pseudofermions} \quad d^\dagger_{j,\itripspin} \ket{0} = \ket{j;4f^2 \; \Gamma_5, \itripspin},\\
    \hat{Q}_j &= \sum_{\ipsspin} f^\dagger_{j,\ipsspin} f^\dphant_{j, \ipsspin} 
    + \sum_{\itripspin} d^\dagger_{j, \itripspin} d^\dphant_{j, \itripspin}
    + \sum_{\ichan} a^\dagger_{j,\ichan} a^\dphant_{j,\ichan},
\end{align}
subject to the constraint that the locally conserved auxiliary-particle number $\hat{Q}_j$ be fixed to unity.
Here, $j$ labels the position of a \ce{Pr} ion, $\ipsspin=\pm$, $\itripspin=0,\pm 1$, $\ichan=\pm 1/2$ are the quantum numbers of the respective CEF states.
Written in terms of these operators, the Pr~$4f$ part of the PAM reflects the energies of $4f^2$ and $4f^3$ states
in Fig.~\ref{fig:CF_scheme_hybf_dxy}(a):
\begin{align}
    H_{\ce{Pr} \; 4f} &= E_{4f^2} \sum_{j,\ipsspin} f^\dagger_{j,\ipsspin}f^\dphant_{j,\ipsspin} + (E_{4f^2} + \Delta) \sum_{j,\itripspin} d^\dagger_{j,\itripspin} d^\dphant_{j,\itripspin}\nonumber \\ \label{eq:CEF}
     & + E_{4f^3} \sum_{j,\ichan} a^\dagger_{j,\ichan} a^\dphant_{j,\ichan} + \sum_j \lambda_j (\hat{Q}_j - \mathds{1}),
\end{align}
Energies are measured relative to the Fermi energy $E_F$ and $\lambda_i \rightarrow \infty$ enforces the  constraint $\hat{Q}=\mathds{1}$~\cite{kroha_NCA_1996}.
The first contribution to the hybridization is the $4f^2, \Gamma_3$ isospin  screened by the conduction electrons projected onto the $\Gamma_8$ states, via valence fluctuations with the Pr~$4f^3$ state,
\begin{align}\label{eq:VF0}
    H_{\mathrm{VF}}^{\Gamma_3} &= \frac{V_{0}}{\sqrt{2}} \sum_{j \ipsspin \ichan}\sgn(\ichan) \; a^\dagger_{j,\ichan} f^\dphant_{j,\ipsspin} \psi^\dphant_{j,-\ipsspin, \ichan} + H.c.\\
    \psi_{j,\ipsspin \ichan} &= \sum_{\vk} \sum_{\iorbital \ispin} e^{i\vk\cdot \RR_j} \; \ctrans^{ \iorbital}_{\ipsspin \ichan; \ispin} (\vk) \; c^\iorbital_{ \ispin}(\vk).\label{eq:Wannier}
\end{align}
Note that (modulo adopting $4f^3$ rather than $4f^1$ as the lowest-energy valence excitation), this is analogous to the hybridization used by other authors~\cite{Van_Dyke_2019,patri_emergent_2020}.
Here, $\psi_{j, \ichan\ipsspin}$ represents the Wannier function of the conduction electrons projected onto the set of \mbox{$|\Gamma_{8},\ipsspin \ichan \ra$} states of the \ce{Pr} atom at position $\mathbf{R}_j$, and the sum is over the conduction-electron orbitals $a$. 
These Wannier functions depend on both the quadrupolar isospin $\ipsspin$ and the Kramers index $\ichan$.
The Wannier projection is captured by the $\vk$-dependent form factor $\ctrans^{\iorbital}_{\ipsspin \ichan; \ispin} (\vk) = \sum_{j,i} e^{-i \vk (\vecR_j - \vecr_i)} \braket{\Gamma^8,  j \ipsspin \ichan | i \iorbital \ispin}$, which projects the atomic orbital $a$ from site $\rr_i$ onto the $|\Gamma_{8}, \ipsspin\ichan\ra$ states of a given Pr atom at position $\vecR_j$~\cite{Note3}.

Taking the orbital, energy, and $\vk$ dependence of the hybridization form factors ${\mathbf{\ctrans}^\dphant}^{\iorbital}(\vk)$ into account beyond the standard Anderson model, the local contribution to the $\Gamma_3$ hybridization function using Eq.~\eqref{eq:VF0} reads,
\begin{align}\label{eq:hybfunc}
    \mathbf{\Gamma}^0(\omega) &= |V_{0}|^2 \sum_{k \iorbital}
        {\mathbf{\ctrans}^\dphant}^{\iorbital}(\vk) \,
        \mathrm{Im} \{\mathbf{G}^{c,0}_{\mathrm{eff}}\}(\omega, \vk) \,
        {\mathbf{\ctrans}^\dagger}^{\iorbital}(\vk)\, , 
\end{align}
where the term under the sum is understood as a matrix product in all the relevant orbital indices.
It incorporates the projection of all bands onto the relevant orbitals, where $\mathbf{G}^{c,0}_{\mathrm{eff}}$ represents a weighted sum of the Green functions over the conduction bands as defined below. The form-factor $\mathbf{\ctrans}(\vk)$ defined in Eq.~\eqref{eq:Wannier} includes the $\vk$-dependent projection of those Wannier states onto the local \ce{Pr} $\Gamma_8$ states. This non-trivial $\vk$ dependence drastically alters the DMFT procedure, where it manifests as a $\vk$-dependent hybridization. For a more detailed discussion, see~\cite{supplement}. The trace of $\mathbf{\Gamma}^0(\omega)$ is plotted in Fig.~\ref{fig:CF_scheme_hybf_dxy}(c).

The second contribution to the hybridization is the $4f^2$, $\Gamma_5$ triplet.
We couple the $\Gamma_5$ operators $d_{j,m}$ to a linear superposition of isospins so as not to break the $\ipsspin$-degeneracy guaranteed by the CEF symmetry~\cite{Note4},
\begin{align}\label{eq:VF1}
    H_{\mathrm{VF}}^{\Gamma_5} = V_1 \sum_{j \ipsspin \itripspin \ichan \ichan^\prime} \Xi_{\itripspin \ichan; \ipsspin \ichan^\prime} a^\dagger_{j,\ichan} d^\dphant_{j,\itripspin} \psi^\dphant_{j,\ipsspin \ichan^\prime} + H.c.
\end{align}
Details of the hybridization matrix $\Xi$ are discussed in the Supplementary Material~\cite{supplement}.
The resulting hybridization function is given analogously to Eq.~\eqref{eq:hybfunc} with $\Xi$ incorporated and a prefactor of $V_1$ instead of $V_0$.
While it is very difficult to compute the values of the hybridization strength $V_1$ from first principles, for concreteness, we set $V_1 = V_0$.

The projection of conduction electrons onto the local $\Gamma_8$ states generally contains contributions from multiple conduction bands, originating from different orbital degrees of freedom labeled by $a$ in Eq.~\eqref{eq:Wannier}.
Previous treatments of \ce{PrV2Al20} have used geometric arguments to select the relevant conduction bands~\cite{Van_Dyke_2019,patri_emergent_2020,han_non-fermi_2022}.
Our DFT analysis shows that mostly aluminum $p$ orbitals and vanadium $d$ orbitals hybridize with Pr $4f$-states, and, given the stoichiometric composition, there are many (140, not counting spin) of such conduction orbitals per unit cell. A full treatment of the orbital-dependent projection is numerically challenging. We therefore justify the following approximations, which significantly improve the feasibility of solving the model.\\
(1) The conduction-electron density of states (DOS) near the Fermi level is dominated by V~$3d$ electrons, as shown by DFT calculations, see Fig.~\ref{fig:CF_scheme_hybf_dxy}(b). This motivates us to ignore the hybridization with Al states, which reduces the number of orbitals considerably.\\
(2) The CEF effect on V sites, with the local $D_{3d}$ point-group symmetry, splits the $d$-electron levels into three irreps: the singlet (ignoring spin) $A_{1g}$ irrep consisting of a $d_{z^2}$ orbital, and two $E_g$ doublets. Of these, the largest contribution to the hybridization function comes from the $\{ \mathrm{d}_{xy}, \mathrm{d}_{x^2 - y^2}\}$ doublet, as can be seen in Fig.~\ref{fig:CF_scheme_hybf_dxy}(c). We thus limit the orbital index $a$ to only run over this doublet, resulting in the total of $N=4\times 2 = 8$ orbitals (because there are 4 V atoms per unit cell).\\
(3) While the DFT operates in the Bloch basis rather than the $N$ vanadium orbitals discussed above, 
for the DMFT calculation we use an effective \ce{V} electron propagator obtained by a sum of projected Bloch bands $n$, where the weight factor 
$w_n^{\iorbital}(\vk)$ projects onto the $\iorbital = \ce{V}d_{xy,x^2-y^2}$ Wannier orbital and the individual Bloch propagators $G_n^{c,0}(\omega, \vk)$ are extracted from DFT. This reduces the DMFT computational effort from treating a large
matrix-propagator at each discretized $\vk$-point to a single-valued one.
In order to extract the unhybridized electron bands entering $G^{c,0}_n(\omega, \vk)$, we need to remove the effect of V-Pr hybridization already present at the DFT level to avoid double-counting. We achieve this by performing the so-called `open core' calculation that places Pr $4f$ electrons into the core, effectively removing their effect on the conduction bands~(see supplementary Section III in \cite{supplement}).

\begin{figure}[tb]
    \centering
    \includegraphics[width=\columnwidth]{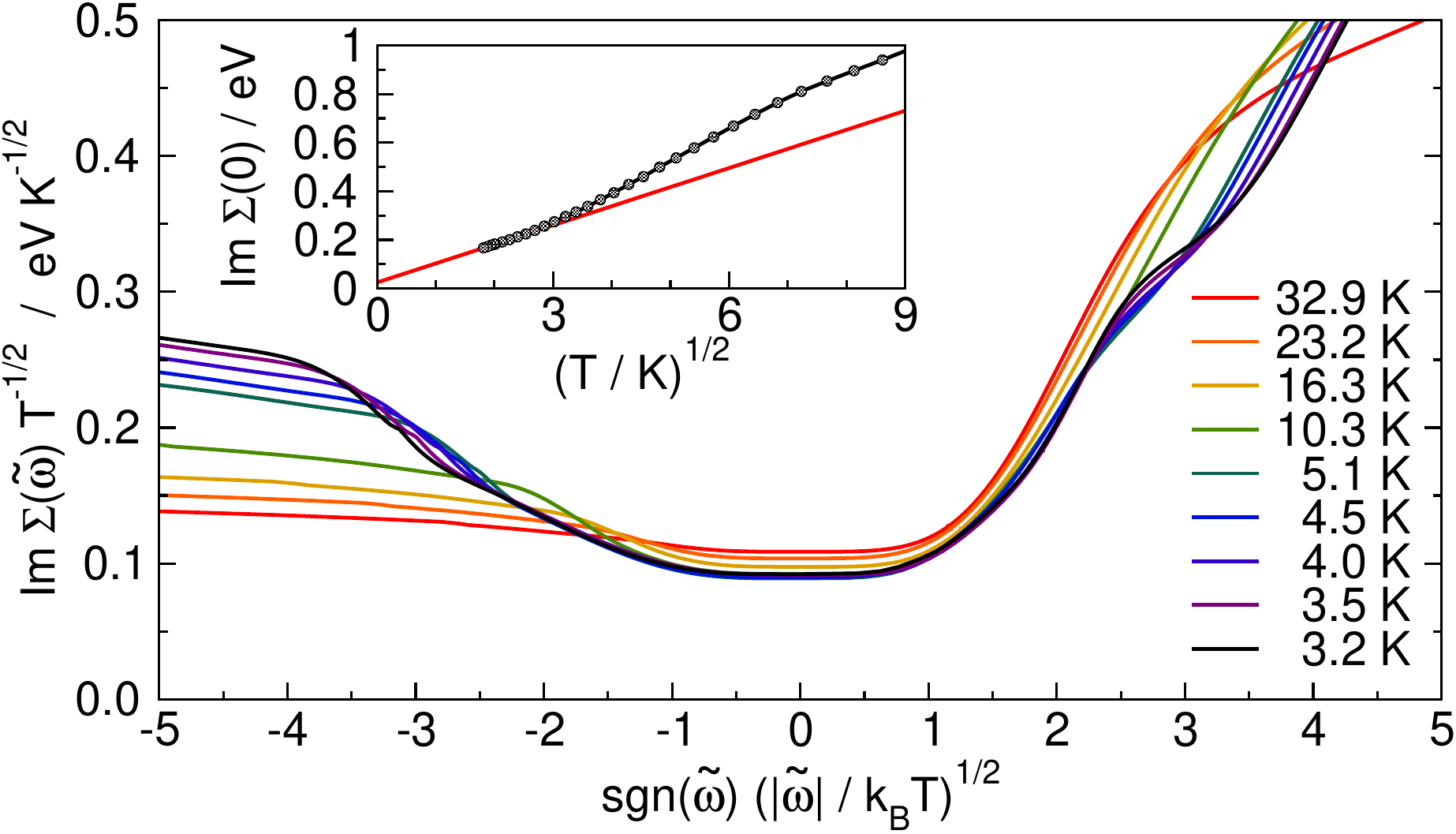}
    \caption{
    Imaginary part of the $\Gamma_3$ contribution to the impurity self-energy divided by $\sqrt{T}$ plotted against the expected 2CK universal scaling function $\sqrt{\tilde{\omega} / k_B T}$ for different temperatures.
    The inset shows $\im \Sigma(\Tilde{\omega}=0)$ plotted against the expected $\sqrt{T}$ dependency and a line as a guide to the eye for the asymptotic behavior.}
    \label{fig:selfenergy}
\end{figure}

\textit{Results.}
 -- The lattice system is solved via DMFT~\cite{georges1992, georges1996, kuramoto_1987}, utilizing the non-crossing approximation (NCA)~\cite{kuramoto1983, keiter_kimball_1971,kroha_NCA_1996} as the impurity solver.  
 The NCA is known to correctly describe the low-energy scales, frequency dependence, and infrared exponents in the 2CK case~\cite{cox_1993}. It is capable of efficiently treating complex multi-orbital systems with satisfactory accuracy while reaching temperatures well below $T_K$~\cite{Ehm2007}.
For the numerical evaluations we used $V_0^2 = V_1^2 = \SI{1.5}~\mathrm{eV}^2$ and a bare hybridization such that the
renormalized CEF splitting of $\Delta_{\mathrm{eff}} / k_B \approx \SI{40}{\kelvin}$, in agreement with experiment.

The spectral function $A(\vk,\omega) = \frac{1}{\pi}\mathrm{Tr}[\mathrm{Im} \mathbf{G}(\vk,\omega)]$, extracted from DMFT, is shown in Fig.~\ref{fig:spectrum}. 
The accumulation of spectral weight near $E_F$ is observable already at $T_{K}^{(0)} \approx \SI{50}{\kelvin}$ which marks the perturbative onset of the Kondo effect, involving both $\Gamma_3$ and $\Gamma_5$ multiplets. Their CEF splitting into Stokes and anti-Stokes satellites~\cite{Ehm2007} at $\omega\approx -\SI{3.2}{\milli\eV}$ and $\omega\approx +\SI{2}{\milli\eV}$ [arrows in Fig.~\ref{fig:spectrum}(c)] 
becomes discernible below $T\approx 20$~K, defining the (renormalized) CEF energy $\Delta_{\mathrm{eff}}$. Note that in Pr the $4f^2 \Gamma_5 \to 4f^3 \to 4f^2 \Gamma_3$ transition process is virtual only, so that the anti-Stokes CEF resonance at $+\SI{2}{\milli\eV}$ appears merely as a shoulder in Fig.~\ref{fig:spectrum}(c)~\cite{Ehm2007}.  
Finally, the non-Fermi liquid shape of the 2CK Abrikosov-Suhl resonance starts to emerge below $T_K\approx \SI{5}{\kelvin}$, which we identify as the 2CK strong-coupling Kondo scale $T_{\mathrm{K}}$.  
\begin{figure}[tb]
    \centering
    \includegraphics[width=\columnwidth]{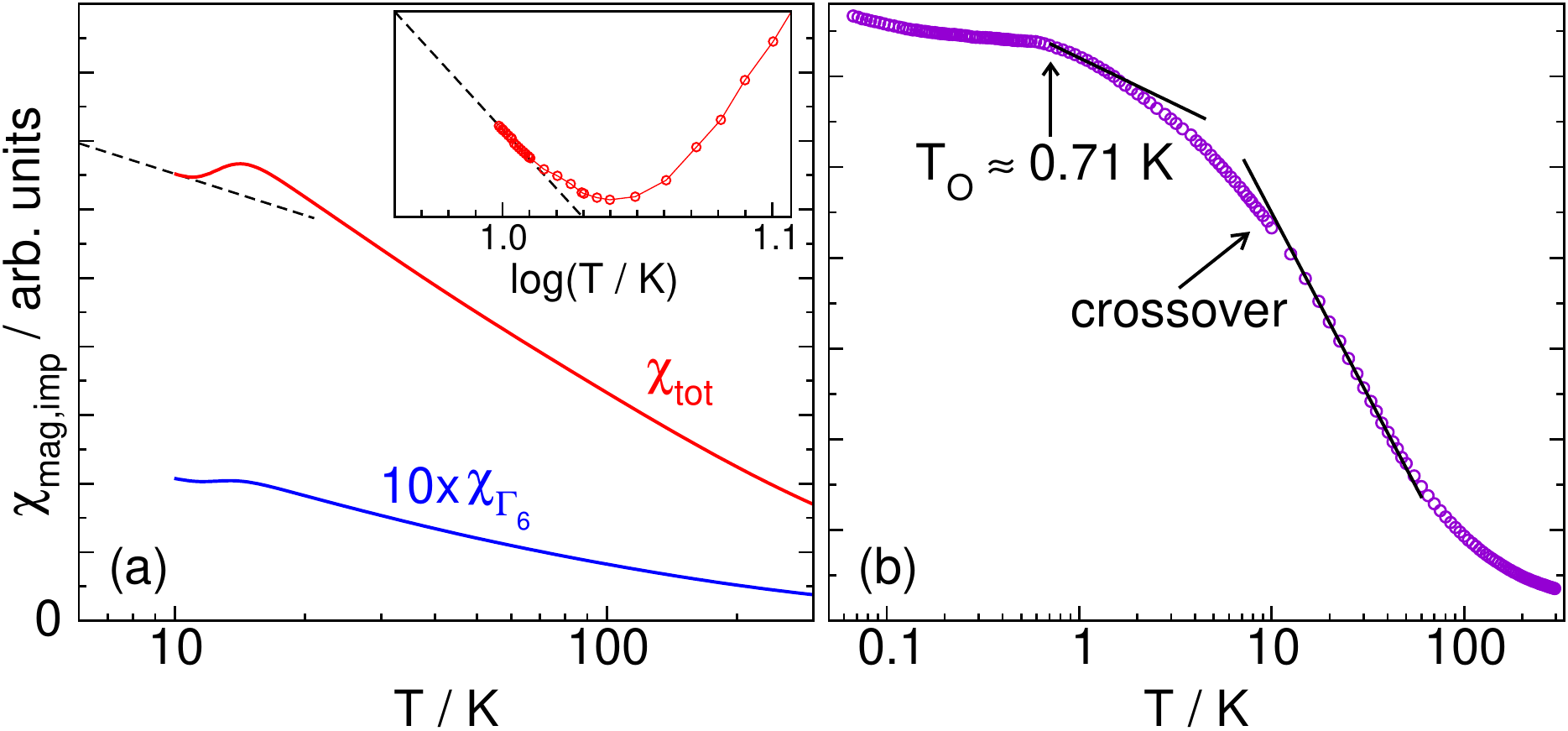}
    \caption{
    Magnetic impurity susceptibility from DMFT. (a) The red curve represents the total susceptibility. The blue curve shows the contribution of the $4f^3$ $\Gamma_6$ doublet (i.e., without the $\Gamma_5$ contribution), multiplied by $10$ for better visibility. The inset shows a blow-up of the low-temperature region. The dashed line is an extrapolation by fitting logarithmic behavior to the low-temperature results.
    (b) Experimental data extracted from Ref.~\cite{araki2014}. The solid lines are linear guides to the eye.
    }
    \label{fig:susceptibility}
\end{figure}
This is confirmed by the imaginary part of the $\Gamma_3$ component of the self-energy $\textrm{Im} \Sigma(\tomega)$ (quasiparticle decay rate) shown in Fig.~\ref{fig:selfenergy} for a sequence of temperatures, where $\tomega = \omega - \omega_0$~\cite{Note5} is the frequency relative to its minimum near $E_{\mathrm{F}}$. 
For $T<T_K\approx \SI{5}{\kelvin}$ it approaches the universal scaling form~\cite{affleck1993, ralph1994}
\beq
\mathrm{Im}\,\Sigma(\tomega)\sim \sqrt{T/T_K} \; F\left(x=\tomega / k_B T\right),
\eeq
where $F(x) \propto \sqrt{x}$ for $\sqrt{x} > 1$ in the $T/T_K \rightarrow 0$ case. 
This is consistent with the experimentally observed temperature dependence of the $\ce{Pr}\,4f$-contribution to electrical resistivity~\cite{sakai2011}.
Note that the lattice model obeys the same scaling behavior as the impurity model \cite{affleck1993}\footnote{%
This has been explored on several occasions for the single-channel case, e.g.~\cite{reinert2001,Ehm2007}, and was also observed in the two-channel case for a wide temperature range in \cite{tsuruta_non-fermi_2015, inui_channel-selective_2020}. However, when non-local effects dominate, e.g.~in an ordered phase, the behavior can deviate.
} %
since the k-integrated spectral density of the Kondo flat band resembles that of the single-impurity case. Around $T=\SI{5}{\kelvin}$, a shoulder at $x \approx 3$ appears and can be attributed to the presence of the $\Gamma_5$ triplet. The shape of $\Sigma(\tomega)$ matches also qualitatively the numerical renormalization group results for a generic 2CK model~\cite{anders_2005}. 

The magnetic susceptibility $\chi(T)$ is the key quantity to compare theory and experiment and to interpret the experimental results on \ce{PrV2Al20}. 
Experiments~\cite{sakai2011, araki2014} reported non-Fermi liquid behavior, $[\chi(T)-\chi(0)]\sim - \sqrt{T}$ seemingly in accordance with the theoretically conjectured temperature dependence of the van Vleck susceptibility by Cox and Makivic~\cite{cox_phenomenology_1994}, at intermediate temperatures, $\SI{2}{\kelvin} \lesssim T \lesssim \SI{30}{\kelvin}$ where, however, a precise functional dependence is hard to extract from the finite temperature range, and which is at odds with the log divergence expected for a 2CK system~\cite{affleck1993}. 
Theoretically, we calculated $\chi(T)$ within DMFT in a small magnetic field $B$ coupled to the conduction spins and all the magnetic local Pr states, numerically taking the derivative with respect to $B$. This includes quantum and thermal fluctuations on the level of DMFT.
The theoretical results, displayed in Fig.~\ref{fig:susceptibility}(a), show that at all temperatures, the total susceptibility $\chi_{\mathrm{tot}}$ is dominated by the magnetic $4f^2~\Gamma_5$ CEF excitation. 
This is expected since the ground-state $\Gamma_3$ doublet, being non-magnetic, provides only a second-order van Vleck susceptibility, and the $4f^3~\Gamma_6$ doublet is only a virtually occupied state. 
All contributions show a logarithmic increase down to $T\approx 20~\mathrm{K}$. By comparison with the spectra of Fig.~\ref{fig:spectrum}(c) we identify this as the perturbative onset~\footnote{
The perturbative onset of the Kondo effect involving both considered CEF multiplets at intermediate temperatures generically results in an initially logarithmic increase of the susceptibility and is \emph{not} a signature unique to two-channel models.
} of the Kondo effect which occurs in the 2CK symmetry-breaking $\Gamma_5$ triplet as well as in the channel-conserving $\Gamma_6$ doublet. 
Below about $T \approx \SI{11}{K}$, separated by an intermediate maximum which may be attributed to the freezing-out of the $\Gamma_5$ CEF excitation, we find a second log increase of $\chi(T)$. 
Since at these low $T$, the conduction-electron channel (magnetic spin) is conserved by the $4f^2~\Gamma_3$ and $4f^3~\Gamma_6$ transitions, this may be understood as the log divergence characteristic of the 2CK effect~\cite{affleck1993}. 
Comparing now these findings with the experimental results of Ref.~\onlinecite{araki2014} shown in Fig.~\ref{fig:susceptibility}(b), 
all the described features can be identified in the experiments, including the high- and low-$T$ log behaviors, where the former is terminated by a crossover point near $\SI{10}{\kelvin}$ that possibly corresponds to the $\Gamma_5$ freeze-out. Intriguingly, the 2CK low-$T$ log behavior seemingly persists through $T_O$.

\textit{Conclusion.} -- 
The rich physics of the 1-2-20 family of Pr-based heavy-fermion materials has provided a canvas for several intriguing observations. 
In the present work, we deliberately ignored the quadrupolar (and possibly octupolar) ordering observed below $T_O\approx 0.71$~K in \ce{PrV2Al20}\cite{sakai2011, ye_measurement_2024}, because this phase has been studied extensively in other works~\cite{hattori2014, ishitobi_triple-_2021, freyer_two-stage_2018, freyer_thermal_2020, patri_emergent_2020}.
Instead, we focused on the non-Fermi liquid strange-metal aspect of these materials, whose interpretation in terms of quantum-critical fluctuations near a quantum-critical point is problematic as it does not appear to require fine-tuning $T_O\to 0$. 
Our key conclusion is that instead, the observed non-Fermi liquid features naturally stem from the two-channel nature of the quadrupolar Kondo effect. 
We showed that the qualitative shape of the magnetic susceptibility is faithfully reproduced on all temperature scales above $T_O$ only by including the excited, channel-symmetry-breaking $\Gamma_5$ CEF states, which, in turn, mask the pure two-channel Kondo behavior at elevated temperatures. The non-Fermi liquid behavior is clearly visible in the self-energy which below \SI{5}{\kelvin} approaches the $\sqrt{\omega/T}$ scaling behavior expected of the two-channel Kondo model in the absence of excited $\Gamma_5$ hybridization. We thus extract $T_K\approx \SI{5}{\kelvin}$ as the two-channel Kondo scale in \ce{PrV2Al20}, which differs from the perturbative Kondo scale $T_K^{(0)} \approx \SI{50}{\kelvin}$. 
We conclude that the two-channel Kondo scenario with crystal-field excitations correctly describes the susceptibility in accordance with experiments over several different temperature regimes. This provides strong support for this scenario to be realized in \ce{PrV2Al20}.
Our findings open up several avenues for future research, including comparing computed spectral functions with future photoemission experiments.
Possible future extension of the theory include a detailed calculation and analysis of the specific heat including the $\Gamma_5$ CEF state and investigation of the magnetic susceptibility below $T_O$.

\textit{Acknowledgements.} The authors thank Y.-B. Kim, R. Flint, and T. Onimaru for discussions. The work at Rice University (F.G. and A.H.N.) was supported by the U.S. National Science Foundation Division of Materials Research under the Award DMR-1917511. The work at the University of Bonn (M.L. and J.K.) was supported in part by the Deutsche Forschungsgemeinschaft (DFG, German Research Foundation) under Germany’s Excellence Strategy – Cluster of Excellence ``Matter and Light for Quantum Computing", ML4Q (390534769), and through the DFG Collaborative Research Center CRC 185 OSCAR (277625399). A.H.N. acknowledges the hospitality of the Aspen Center for Physics,  supported by the National Science Foundation grant PHY-1607611,  where this collaboration was initiated.

%

\clearpage
\setcounter{equation}{0}
\setcounter{figure}{0}
\setcounter{table}{0}

\title{Supplemental Material for \texorpdfstring{\\ ``\thetitle"}{``\thetitle"}}

\maketitle
\renewcommand{\theequation}{S\arabic{equation}}
\renewcommand{\thefigure}{S\arabic{figure}}
\renewcommand{\bibnumfmt}[1]{[S#1]}
\renewcommand{\citenumfont}[1]{S#1}
%
%
\section{DFT Results and Crystal Structure}

\begin{figure}[h]
    \centering
    \includegraphics[width=0.94\columnwidth]{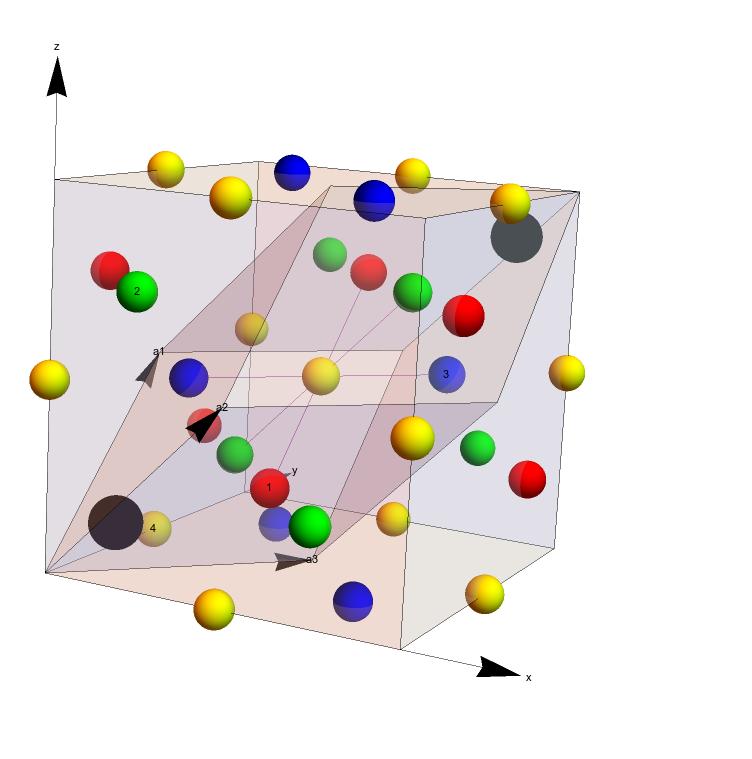}
    \vspace{-5mm}
    \caption{
        Primitive unit cell of \ce{PrV2Al20} illustrated with VESTA \cite{supp_momma_vesta_2011}. Black balls represent \ce{Pr} atoms, the other colors represent different classes of \ce{V} atoms.}
    \label{fig:prv2al20_primunitcell}
\end{figure}

The crystal structure of \ce{PrV2Al20} is face-centered cubic (fcc), with cell length $a=b=c=14.567(3)$\AA, angle $\ipsspin=\itripspin=\gamma=90^{\circ}$. The space group name is Fd-3m. The cell formula unit is $Z=8$, and there are eight \ce{Pr} atoms and sixteen \ce{V} atoms per conventional unit cell. A primitive unit cell contains two \ce{Pr} and four \ce{V} atoms, as shown in Fig.~\ref{fig:prv2al20_primunitcell}. \ce{Pr} atoms form a diamond lattice, while \ce{V} atoms form four distinct sets of fcc structures. In Fig.~\ref{fig:prv2al20_primunitcell}, V atoms are represented by balls of four different colors, each color forms an fcc structure. While sixteen \ce{Al} atoms form Frank-Kasper cages around each \ce{Pr} \cite{supp_onimaru2016}, twelve \ce{V} atoms (three of each color) also form a regular structure around each \ce{Pr} atom. Due to the location of \ce{Pr} close to a corner, this is not directly clear from just the primitive unit cell. Of those six \ce{V} atoms, only three sit in the same unit cell as the \ce{Pr} atom. Note that atoms on edges and surfaces of the unit cell have quarter and half weights in this counting.

%
%
\section{Details of the Structure factor}
The effective conduction electrons in our model are given by a projection of all bands onto \ce{V} $d$ orbitals. The hybridization is, however, mediated by those electrons projected onto $\Gamma_8$ states at the location of the \ce{Pr} atom. For this, the overlap of the orbital Wannier functions at each \ce{V} site has to be calculated, resulting in a $\vk$-dependent structure factor $\mathbf{\ctrans}(\vk)$.

The real-space hybridization structure factor is
\begin{equation} \label{eq:hybft1}
    \ctrans_{\ipsspin \ichan; \ispin}^{\iorbital} (\vecR_j - \vecr_i)
    = \braket{\ce{Pr}~\Gamma^8;  j \ipsspin \ichan |  \ce{V}; i \iorbital \ispin} .
\end{equation}
Here, $j$ labels the \ce{Pr} position $\vecR_j$ and $i$ the \ce{V} position $\vecr_i$. Additionally, $\ipsspin = \pm$ and $\ichan = \uparrow, \downarrow$ are the $\Gamma_8$ quantum numbers, $\iorbital$ and $\ispin$ represent \ce{V} orbital and spin quantum numbers. Note that the structure factor only depends on the distance $\vecR_j - \vecr_i$ and it's Fourier transform is diagonal in momentum space.
The structure factor in $\vk$-space,
\begin{align}\label{eq:struct_fourier}
    \ctrans_{\ipsspin \ichan; \ispin}^{\iorbital}(\vk) &= \sum_{i,j} e^{-i\vk(\vecR_j - \vecr_j)} \, \ctrans_{\ipsspin \ichan; \ispin}^{\iorbital} (\vecR_j - \vecr_i)\, ,
\end{align}
does not depend on $\vecR_j$ anymore. The \ce{Pr} position is however explicitly treated in the projection, which reads,
\begin{align}
    \psi_{j,\ipsspin \ichan} &= \sum_{i}\sum_{\iorbital, \ispin} \ctrans_{\ipsspin \ichan; \ispin}^{\iorbital} (\vecR_j - \vecr_i) \, c_{\ispin}^{\iorbital}(\vecr_i)\\
    &= \sum_{\vk, \vk^\prime, i}\sum_{\iorbital, \ispin} e^{i\vk(\vecR_j - \vecr_i)} \, \ctrans_{\ipsspin \ichan; \ispin}^{\iorbital}(\vk) \, e^{i\vk^\prime \vecr_i} \, c_{\ispin}^{\iorbital}(\vk^\prime)\\
    &= \sum_{\vk} \sum_{\iorbital, \ispin} e^{i \vk \vecR_j} \, \ctrans_{\ipsspin \ichan; \ispin}^{\iorbital}(\vk) \, c_{\ispin}^{\iorbital}(\vk).
\end{align}
For convenience, we define a matrix $\mathbf{\ctrans}^{\iorbital}_j(\vk)$ as
\begin{align}
    \left[\mathbf{\ctrans}^{\iorbital}_j (\vk)\right]_{\ipsspin \ichan ; \ispin} 
    &:= \ctrans_{j \ipsspin \ichan; \ispin}^{\iorbital}(\vk)
    := e^{i \vk \vecR_j} \, \ctrans_{\ipsspin \ichan; \ispin}^{\iorbital}(\vk) \,.
\end{align}
Only two orbitals $\iorbital \in \{ \mathrm{d}_{xy}, \mathrm{d}_{x^2 - y^2}\} =: \mathcal{D}$ are considered, as discussed in the main text. Due to the local nature of DMFT, we only need to consider $\vecR_j$ in a single unit cell, and due to the two \ce{Pr} atoms being crystallographically identical, we only take one of them ($\vecR_1$) into account. Additionally, we are unable to resolve the two degenerate orbitals and the individual \ce{V} sub-lattices within DFT and, therefore, sum over them.
In order to numerically calculate this $\vk$-dependent structure factor, the position vectors of all structure-factor relevant atoms are needed.

As shown in Fig.~\ref{fig:prv2al20_primunitcell}, There are four \ce{V} atoms per primitive unit cell. The lattice basis vectors are
\begin{align}
    \mathbf{a}_1 &= \frac{1}{2}\begin{pmatrix}0\\a\\a\end{pmatrix}, &
    \mathbf{a}_2 &= \frac{1}{2}\begin{pmatrix}a\\0\\a\end{pmatrix}, &
    \mathbf{a}_3 &= \frac{1}{2}\begin{pmatrix}a\\a\\0\end{pmatrix}.
\end{align}
In the basis of those, the coordinates of \ce{V} atoms are
\begin{align*}
    \mathbf{v}_{\mathrm{red}} &= \frac{1}{2}\begin{pmatrix}0\\1\\1\end{pmatrix}, &
    \mathbf{v}_{\mathrm{green}} &= \frac{1}{2}\begin{pmatrix}1\\0\\1\end{pmatrix},\\
    \mathbf{v}_{\mathrm{blue}} &= \frac{1}{2}\begin{pmatrix}1\\1\\0\end{pmatrix}, &
    \mathbf{v}_{\mathrm{yellow}} &= \frac{1}{2}\begin{pmatrix}1\\1\\1\end{pmatrix},
\end{align*}
and those of the two Pr atoms are 
\begin{align*}
    \vecR_1 &= \frac{1}{8}\begin{pmatrix}1\\1\\1\end{pmatrix}, &
    \vecR_2 &= \frac{1}{8}\begin{pmatrix}7\\7\\7\end{pmatrix}.
\end{align*}
As explained before, we only take the \ce{Pr} atom at $\vecR_1$ into account. It has twelve neighboring \ce{V} atoms, three \ce{V} atoms for each fcc sub-lattice $l \in \{1,2,3,4\}$ or equivalently $\{\mathrm{Red},\mathrm{Green},\mathrm{Blue},\mathrm{Yellow}\}$. Each of those three \ce{V} atoms is then labeled by $l_s \in \{1,2,3\}$. Their position vectors $\mathbf{r}_{l,l_s}$ are listed below.
\begin{align*}
    \mathbf{r}_{1,1} &= \frac{1}{2}\begin{pmatrix}0\\1\\1\end{pmatrix} &
    \mathbf{r}_{1,2} &= \frac{1}{2}\begin{pmatrix}0\\1\\-1\end{pmatrix} &
    \mathbf{r}_{1,3} &= \frac{1}{2}\begin{pmatrix}0\\-1\\1\end{pmatrix} \\
    \mathbf{r}_{2,1} &= \frac{1}{2}\begin{pmatrix}1\\0\\1\end{pmatrix} & 
    \mathbf{r}_{2,2} &= \frac{1}{2}\begin{pmatrix}1\\0\\-1\end{pmatrix} &
    \mathbf{r}_{2,3} &=\frac{1}{2}\begin{pmatrix}-1\\0\\1\end{pmatrix} \\
    \mathbf{r}_{3,1} &= \frac{1}{2}\begin{pmatrix}1\\1\\0\end{pmatrix} & 
    \mathbf{r}_{3,2} &= \frac{1}{2}\begin{pmatrix}1\\-1\\0\end{pmatrix} &
    \mathbf{r}_{3,3} &= \frac{1}{2}\begin{pmatrix}-1\\1\\0\end{pmatrix} \\
    \mathbf{r}_{4,1} &= \frac{1}{2}\begin{pmatrix}-1\\1\\1\end{pmatrix} & 
    \mathbf{r}_{4,2} &= \frac{1}{2}\begin{pmatrix}1\\-1\\1\end{pmatrix} &
    \mathbf{r}_{4,3} &= \frac{1}{2}\begin{pmatrix}1\\1\\-1\end{pmatrix}
\end{align*}
We only consider the contribution of those twelve atoms, them being the nearest neighbors of \ce{Pr}. 

Collecting everything above, the final expression for the structure factor reads,
\begin{align}
    \left[\mathbf{\ctrans}_j (\vk)\right]_{\ipsspin \ichan ; \ispin} 
    &= e^{i\vk \vecR_j} \sum_{\iorbital \in \mathcal{D}} \ctrans_{\ipsspin \ichan; \ispin}^{\iorbital}(\vk),\\
    \ctrans_{\ipsspin \ichan; \ispin}^{\iorbital}(\vk)
    &= \sum_{l, l_s} e^{-i\vk(\vecR_1 - \vecr_{l,l_s})} \ctrans_{\ipsspin \ichan; \ispin}^{\iorbital} (\vecR_j - \vecr_{l,l_s}),
\end{align}
which is the structure factor we use in our calculation. Note that the second line is just the Fourier transform in the reduced set of contributing \ce{V} atoms, projected onto a single \ce{Pr} atom.
%
%
\section{Effective Conduction-Electron Hamiltonian}

\subsection{Isolating the conduction bands from Pr \texorpdfstring{$4f$}{4f} core levels}\label{sec:open-core}

In order to formulate the effective model for PrV$_2$Al$_{20}$ -- the periodic Anderson model -- one needs to isolate the conduction bands from the localized 4f electrons of Pr. There are two issues that one encounters on this path: (i) the \textit{ab initio} DFT severely exaggerates the 4f electron bandwidth, not able to capture the localizing effect of strong electron correlations; and (ii) the problem of ``double counting'', namely that the effect of $c$-$f$ hybridization is already captured to some extent by DFT. The latter issue, in particular, is troublesome, as seen in DFT+U and DFT+DMFT attempts to capture the correlated problem.

In order to overcome these issues, we adopt the ``open core'' method, developed by Peter Blaha as part of the Wien2k package~\cite{supp_wien2k} in order to rectify the shortcoming of DFT that fails to localize the $4f$ electrons. In its essence, the method ``freezes'' the Pr $4f$ electrons into the core so as to enforce the Pr$^{3+}$ electron configuration [Xe]$4f^2$ with precisely two electrons in the 4f shell. (Parenthetically, it might be more appropriate to call this a ``frozen core'' method -- the name ``open core'' is historical).
In order to achieve this, the atomic potential is artificially shifted downwards within the atomic sphere so as to ensure that the $4f$ states have a bound, negative-energy eigenvalue. This effectively removes the $4f$ electrons from the valence and hence completely removes their hybridization with the conduction electrons near the Fermi level. As a result, this also negates the issue with double-counting, which is ideal for the subsequent DMFT treatment of the periodic Anderson model.

We hasten to add that the effect of ``freezing'' the $4f^2$ configuration results in the side effect of shifting the chemical potential relative to its value in the all-electron DFT calculation. This is to be expected since the conduction electron bands redistribute once the hybridization with the $4f$ levels of Pr has been ``switched off.''
We, therefore, adjust the chemical potential in all subsequent calculations so that the features of the density of states away from the Fermi level become properly aligned with those from the all-electron DFT. This method of ``realigning'' the chemical potential is well established in the open core calculations~\cite{supp_wien2k}.

\subsection{Propagators for the conduction bands}
The output of the aforementioned ``open core'' DFT calculation is a set of conduction bands without the 4f levels of praseodymium, encoded in the Kohn--Sham energy eigenvalues $\epsilon^n_{\vk}$ for the $n^\text{th}$ band at a given $\vk$ point. 
The Hamiltonian is, therefore, quadratic and given already in an energy eigenbasis
\begin{align}
    H_{\mathrm{DFT}} &= \sum_{n,\ispin,\vk} \epsilon^n_{\vk} c^\dagger_{n \ispin \vk} c^{\dphant}_{n \ispin \vk} \, .
\end{align}
A density of states (DOS) can be obtained for each individual band by
\begin{align}
    \rho^n(\omega) &= \sum_{\vk} \delta(\omega - \epsilon^n_{\vk}) = \lim_{\eta \rightarrow 0^+} \im \sum_{\vk} \frac{\pi^{-1}}{\omega - \epsilon^n_{\vk} - i \eta} \, .
\end{align}
Equally, the imaginary part of the advanced Green function $G^A_n(\omega)$ is $\pi \rho^n(\omega)$. Since localized atomic orbitals are typically not eigenstates of the crystal, a projection onto such orbitals involves contributions from all bands. An effective, free propagator for a specific orbital $\iorbital$ is then given by a weighted sum
\begin{align}
    G^{c,0}_{\mathrm{eff}}(\omega, \vk) = \sum_n w_n^{\iorbital}(\vk) \, G_n^{c,0}(\omega, \vk),
\end{align}
where the $\vk$-dependent weight factors $w_n^{\iorbital}(\vk)$ and band energies $\epsilon^n_{\vk}$, and therefore also $G_n^{c,0}(\omega, \vk)$, are calculated via DFT. On the level of an action $S$, one can then just write the action for an effective model as
\begin{align}
    S^c_{\mathrm{eff}} &= \sum_{i \omega_m, \vk, \ispin}
    \bar{c}_{m \vk \ispin} G_{\mathrm{eff}}^{-1}(i \omega_m, \vk) c_{m \vk \ispin} .
\end{align}
Although everything can be formulated in terms of path integrals, we prefer a Hamiltonian formulation. Due to the weighted sum of propagators, this is not done in a straightforward way. Technically, it can be back-engineered from the Lagrangian density. In the DMFT formulation, however, one only needs the inverse Green function of the conduction electrons and not the Hamiltonian itself. 
%
%
\section{Dominant valence fluctuating channel in \texorpdfstring{P\titlelowercase{r}V$_2$A\titlelowercase{l}$_{20}$}{PrV2Al20}}

The previous studies of PrV$_2$Al$_{20}$, such as  Ref.~\cite{supp_Van_Dyke_2019}, have tacitly assumed that the dominant fluctuating valence channels are between the non-Kramers $4f^2$ ground state and an excited $4f^1$ state with one electron removed. At the same time, it is not obvious whether the $4f^3$ valence state can be ignored. Here, we use the \textit{ab initio} calculations to determine the dominant valence fluctuating channel.

We employ the above described ``open core'' method (Sec.~\ref{sec:open-core}) to fix the desired number of electrons, be it $n=1$ or $n=3$, in Pr $4f$ shell. By comparing the difference of the resulting eigenvalue of the $4f^n$ core state relative to the $4f^2$ ground state, we found it to be $\Delta E_{2 \leftrightarrow 3} = E_{4f^3} - E_{4f^2} \approx \SI{2.2}{\eV}$. Performing the analogous calculation for the $4f^1$ state, we find a significantly higher excitation energy $\Delta E_{2 \leftrightarrow 1} \approx \SI{3.8}{\eV}$. We, therefore, adopt $4f^2 \leftrightarrow 4f^3$ as the dominant Kondo hybridization channel in what follows.

%
%
\section{Impurity Hamiltonian and Auxiliary Particles}
The local impurity part is modeled by a multi-channel single-impurity Anderson model (SIAM) \cite{supp_schiller_mixed-valent_1998, supp_cox1998}. The Hubbard operators obey a non-canonical commutation relation and are not well suited for perturbative field theory. We therefore resort to the slave-boson representation (c.f. Eqs.~(1)--(3) in the main text) of the Hamiltonian, given by Eqs.~(4)--(6) and (8) in the main text, in which we have to send $\lambda_j \rightarrow \infty$ to enforce the constraint $\hat{Q} = \mathds{1}$, projecting the enlarged Hilbert space to the physical sector. Such a model is the infinite-$U$ limit of a modified SIAM
\begin{align}
    H^{\text{imp}}_{SIAM} &= \sum_{m ,\ipsspin_m, \ichan} \epsilon^{\mathrm{imp}}_m t^\dagger_{m \ipsspin_m \ichan} t^\dphant_{m \ipsspin_m \ichan}\\ \nonumber
    \qquad &+ U \sum_{\substack{m,m^\prime \\ \ipsspin_m \neq \ipsspin_{m^\prime}}}\sum_{\ichan, \ichan^\prime} \hat{n}^{(t)}_{m \ipsspin_m \ichan} \hat{n}^{(t)}_{m^\prime \ipsspin_{m^\prime} \ichan^\prime},
\end{align}
where $m$ labels impurity orbitals with quantum numbers $\ipsspin_m$ and $\ichan$ is a conserved hybridization channel. The usual convention for the field operators would be to use $d$ or $f$, but since both are already used for the pseudofermions, we resort to using $t$. Under $U \rightarrow \infty$, the impurity field operator can be written as $t^\dagger_{m \ipsspin_m \ichan} = f^\dagger_{m \ipsspin_m} b_{\bar{\ichan}}$, where $\bar{\ichan} = -\ichan$ is the conjugate channel. Since we are considering the $4f^3$ state as the first excited one, one has to construct a hole model with infinite hole-repulsion $U^\prime$ to get analogous expressions. One can, therefore, define impurity propagators in a meaningful way in analogy to the regular infinite-$U$ SIAM or, in the lattice case, the periodic Anderson impurity model, ultimately allowing for a standard DMFT derivation.
%
%
\section{Relevant CEF multiplets in \texorpdfstring{P\titlelowercase{r}V$_2$A\titlelowercase{l}$_{20}$}{PrV2Al20}}
The experimentally determined CEF ground state of the Pr$^{2+}$ ion is the $\Gamma_3$ doublet of the cubic $T_d$ point group. Hund's rules for the $4f^2$ configuration state that the spin-orbit coupled ground state is a $J=4$ state, whose $m_J$ states provide a basis in which the CEF states can be expressed \cite{supp_kusunose_interplay_2005}.

The index $\ipsspin=\pm$, dubbed isospin, enumerates the non-Kramers $|4f^2, \Gamma_3\ipsspin\ra$ states:
\begin{align}
    \ket{\Gamma_3-} &= 
    \frac{1}{\sqrt{2}}\left(\ket{+2} + \ket{-2}\right)\label{eq:Gamma3}  \\
    \ket{\Gamma_3+} &= \frac{1}{2\sqrt{6}} \left(\sqrt{7} \left[ \ket{+4} + \ket{-4} \right] - \sqrt{10}\ket{0} \right).
\end{align}
Additionally, first excited CEF state is the $\Gamma_5$ triplet, whose states $\ket{4f^2 \; \Gamma_5, \itripspin}$ with $\itripspin \in \{-1,0,1\}$ are
\begin{align}
    \ket{\Gamma_5, \pm} &= \pm \frac{1}{4 \sqrt{2}}\left( \sqrt{7} \ket{\pm 3} - \ket{\mp 1} \right)\\
    \ket{\Gamma_5, 0} &= \frac{1}{\sqrt{2}} \Big( \ket{+2} - \ket{-2} \Big).
\end{align}

In the same manner, the $4f^1$ states mediating the valence fluctuations can be expressed in their Hund's rule ground state of $J=5/2$. Here, we only consider The $\Gamma_8$ quartet, which is given by
\begin{align}
    \ket{\Gamma_8^{+ \sfrac{\uparrow}{\downarrow}}} &= \frac{1}{\sqrt{6}} \left(\sqrt{5}\ket{\pm\sfrac{5}{2}} + \ket{\mp\sfrac{3}{2}} \right)\\
    \ket{\Gamma_8^{- \sfrac{\uparrow}{\downarrow}}} &=\ket{\pm\sfrac{1}{2}}.
\end{align}
%
%
\section{CEF Multiplet Hybridization Matrix Elements}\label{sec:cef_hyb}
The CEF ground state of the $4f^2$ transforms as $\Gamma_3$, the ground state of $4f^3$ likely as $\Gamma_6$ irreps of the $T_d$ group. A hybridization term connecting $4f^2$ and $4f^3$ should therefore transform as $\Gamma_6 \otimes \Gamma_3 = \Gamma_8$. Conduction electrons mediating this transition are to be projected onto the local $\Gamma_8$ quartet. Excited CEF levels of $4f^2$ will be frozen out exponentially at low temperatures. One expects this hybridization to be most important for the discussion of the two-channel quadrupolar Kondo physics in \ce{PrV2Al20}. 
As discussed in the main text, however, we also need to include the first excited $4f^2$ CEF level to faithfully investigate the magnetic susceptibility. The 
associated irrep is $\Gamma_5$ and transforms like a spin-triplet, labeled by the index $\itripspin$ in Eqs.~(2)--(4) and (8) in the main text. The corresponding hybridization term transforms as $\Gamma_6 \otimes \Gamma_5 = \Gamma_7 \oplus \Gamma_8$, requiring an additional $\Gamma_7$ doublet on the conduction-electron side. By choosing a basis for $\Gamma_8$ that trivializes the hybridization with $\Gamma_3$ we get a non-trivial hybridization with $\Gamma_5$. This could be mitigated by choosing an appropriate $\Gamma_7 \oplus \Gamma_8$ basis for this hybridization separately at the cost of having two different projected conduction-electron fields: $\psi$ in Eq.~(6) (main text) and $\psi^\prime$ given by a modified $\vk$-dependent structure factor. This increases the DMFT complexity and numerical cost significantly, which is why we chose to stay in the original basis.

In an effort to simplify the model and in the context of the main interest of this Letter, we take the conduction electrons to transform as the $\Gamma_8$ quartet, neglecting the $\Gamma_7$ part on the right-hand side of $\Gamma_6 \otimes \Gamma_5 = \Gamma_7 \oplus \Gamma_8$. Since we describe the symmetries of the local Pr 4f orbitals faithfully, this does not change any of our conclusions.

Once additional CEF levels are taken into account, one needs to carefully consider relative phases and hybridization strengths. The matrix elements we use are calculated from dipole-moment conservation and an appropriate combination of states using Clebsch-Gordan coefficients. For the $\Gamma_3$ ground state, we have local hybridization terms of the structure $f^\dagger_\ipsspin a^\dphant_\ichan \psi^\dagger_{\ipsspin^\prime \ichan^\prime}$. If we take $f^\dagger_\ipsspin$ as the target state, we first need to particle-hole-transform \emph{with spin- and quadrupole-moment flipped} in the conduction electrons $\psi^\dagger_{\ipsspin \ichan} \rightarrow \zeta_{-\ipsspin -\ichan}$, resulting in a hybridization $f^\dagger_\ipsspin a^\dphant_\ichan \zeta^\dphant_{\ipsspin^\prime \ichan^\prime}$. Since $\ipsspin$ is conserved in the scattering, we immediately get $\ipsspin = \ipsspin^\prime$. The dipole moment must vanish in order for it to transform as a scalar, indicating that conduction-electron holes and slave bosons must form a spin singlet. We therefore get
\begin{align}
    H_{VF, \mathrm{loc}}^{\Gamma_3} &= V_0 \sum_{\ipsspin} f^\dagger_\ipsspin \frac{1}{\sqrt{2}} 
    \left( a^\dphant_\uparrow \zeta^\dphant_{\ipsspin \downarrow} - a^\dphant_\downarrow \zeta^\dphant_{\ipsspin \uparrow} \right) + H.c.\\
    &= \frac{V_0}{\sqrt{2}} \sum_{\ipsspin} \mathrm{sgn}(\ichan) f^\dagger_\ipsspin a^\dphant_\ichan \zeta^\dphant_{\ipsspin -\ichan} + H.c.\\
    &= \frac{V_0}{\sqrt{2}} \sum_{\ipsspin} \mathrm{sgn}(\ichan) f^\dagger_\ipsspin a^\dphant_\ichan \psi^\dagger_{-\ipsspin \ichan} + H.c. .
\end{align}
The conjugation (spin- and quadrupole flip) of quantum numbers in the particle-hole transformation was necessary to make sure the Hamiltonian conserves spin and quadrupole moment in both representations. This is needed since we \emph{only transform the conduction electrons} and not the whole Hamiltonian.

For the excited state, which transforms like a spin-triplet, we have to apply the same logic. First, the hybridization term is $d^\dagger_\itripspin a^\dphant_\ichan \psi^\dagger_{\ipsspin \ichan^\prime} \rightarrow d^\dagger_\itripspin a^\dphant_\ichan \zeta^\dphant_{\ipsspin \ichan^\prime}$ with $\psi^\dagger_{\ipsspin \ichan} \rightarrow \zeta_{-\ipsspin -\ichan}$ as before. Here, $\itripspin \in \{-1,0,1\}$. We need to combine the spins of $a$ and $\zeta$ to a triplet. The local hybridization contribution of $\Gamma_5$ to the Hamiltonian is therefore
\begin{align}\nonumber
    H_{VF, \mathrm{loc}}^{\Gamma_5} &= V_1 \sum_{\ipsspin} \eta_\ipsspin \Big[ d^\dagger_0 \frac{1}{\sqrt{2}} 
    \left( a^\dphant_\uparrow \zeta^\dphant_{\ipsspin \downarrow} + a^\dphant_\downarrow \zeta^\dphant_{\ipsspin \uparrow} \right) \\
    &\qquad+ d^\dagger_{+1} a^\dphant_\uparrow \zeta^\dphant_{\ipsspin \uparrow}
    + d^\dagger_{-1} a^\dphant_\downarrow \zeta^\dphant_{\ipsspin \downarrow} \Big] + H.c..
\end{align}
The factor $\eta_\ipsspin$ has to be determined to match the right-side quadrupole moment with the quadrupole moment of $\Gamma_5$. This will generally lead to a superposition. Since it really is a superposition and not a combination of quadrupole moments of two particles, a reasonable approach would be to define $\eta_\ipsspin$ such that $\sum_\ipsspin |\eta_\ipsspin|^2 = 1$. The simplest choice is $\eta_+ = \eta_-$, with which we get $\eta_\ipsspin = 1/\sqrt{2}$, resulting in
\begin{align}\nonumber
    H_{VF, \mathrm{loc}}^{\Gamma_5} &= \frac{V_1}{\sqrt{2}} \sum_{\ipsspin} \Big[ d^\dagger_0 \frac{1}{\sqrt{2}} 
    \left( a^\dphant_\uparrow \zeta^\dphant_{\ipsspin \downarrow} + a^\dphant_\downarrow \zeta^\dphant_{\ipsspin \uparrow} \right) \\
    &\qquad+ d^\dagger_{+1} a^\dphant_\uparrow \zeta^\dphant_{\ipsspin \uparrow}
    + d^\dagger_{-1} a^\dphant_\downarrow \zeta^\dphant_{\ipsspin \downarrow} \Big] + H.c..
\end{align}
With this choice of prefactors, setting $V_1 = V_0$ results in identical local hybridization function strengths for $\Gamma_3$ and $\Gamma_5$. 

Additionally to what was discussed before, one can also modify the prefactors such that the hybridization transforms like a scalar, which can be achieved by understanding the operators $\xi$ and $a$ combined as a spin-1 operator $\zeta$, combining $d$ and the $\zeta$-hole (with flipped spin) $\Tilde{\zeta}$ to a total spin zero such that
\begin{align}\nonumber
    H_{VF, \mathrm{loc}}^{\Gamma_5} &= \frac{V_1}{\sqrt{6}} \sum_{\ipsspin} \eta_\ipsspin \left( d^\dagger_{+1} \Tilde{\zeta}^\dagger_{\ipsspin, -1}
    + d^\dagger_{-1} \Tilde{\zeta}^\dagger_{\ipsspin, +1} - d^\dagger_0 \Tilde{\zeta}^\dagger_{\ipsspin,0}\right)\\ 
    & \qquad + H.c.,
\end{align}
or, substituting everything back,
\begin{align}
    H_{VF, \mathrm{loc}}^{\Gamma_5} &= \frac{V_1}{\sqrt{3}} \sum_\ipsspin \eta_\ipsspin \Big[ d^\dagger_{+1} a^\dphant_\uparrow \psi^\dagger_{\ipsspin \downarrow}
    + d^\dagger_{-1} a^\dphant_\downarrow \psi^\dagger_{\ipsspin \uparrow} \\
    & \qquad - d^\dagger_0 \frac{1}{\sqrt{2}} 
    \left( a^\dphant_\uparrow \psi^\dagger_{\ipsspin \uparrow} + a^\dphant_\downarrow \psi^\dagger_{\ipsspin \downarrow} \right) \Big] + H.c..
\end{align}
When setting $V_1 = V_0 \sqrt{3/2}$, this is identical to the previous case with the exception of a minus sign in front of the $\itripspin = 0$ term, which turns out to be insignificant in our calculation. 
We found that numerically, an imbalance between the two induces a much larger CEF splitting than the experimental results \cite{supp_sakai2011} suggest. This has to be mitigated via a modified bare splitting in the Hamiltonian.

Generally, we could also have an imbalance between the two quadrupole momenta, e.g. $\eta_- = 0$. This is reflected by the fact that $\Gamma_6 \otimes \Gamma_8 = \Gamma_3 \oplus \Gamma_4 \oplus \Gamma_5$. Forming a dipole singlet on the left side leaves a quadrupole moment unpaired, resulting in the $\Gamma_3$ doublet. Conversely, the quadrupole moment is still not paired when forming a spin triplet. Now, the right-hand side needs to have a threefold spin-1 quantum number in addition to the twofold quadrupole orientation. One can imagine the most extreme case in which each quadrupole moment orientation is individually attached to a triplet, resulting in two potentially different triplets, which is reflected in the appearance of the $\Gamma_4$ and $\Gamma_5$ triplets on the right. Only considering one of the triplets potentially induces an imbalance between the two quadrupole moments, resulting in a splitting at all temperatures. Since this is unphysical, we stick to an $\ipsspin$-symmetric $\Gamma_5$ coupling. One could remedy this by also considering $\Gamma_4$, which is beyond the scope of this letter.

%
%
\section{Details of the DMFT solver}
The DMFT construction is the standard construction for a PAM with $\vk$-dependent hybridization \cite{supp_georges1996} but with the aforementioned free, effective conduction-electron propagator. This leads to the DMFT lattice Green functions
\begin{align}
    \mathbf{G}^c (i \omega_n) &= \sum_{\vk} \frac{1}{\{\mathbf{G}^{0}_{\text{eff}}(i\omega_n, \vk)\} ^{-1}- \mathbf{\Delta}^c(i \omega_n, \vk)}\\
    \mathbf{G}^{\text{imp}}(i \omega_n) &= \frac{1}{i \omega_n - \mathbf{\epsilon^{\text{imp}}} - \mathbf{\Delta}^{\text{imp}}(i \omega_n) - \mathbf{\Sigma}(i \omega_n) },
\end{align}
with 
\begin{align}
    \mathbf{\epsilon^{\text{imp}}} &= \epsilon \mathds{1}_{\Gamma_3 \otimes \Gamma_6} \oplus (\epsilon + \Delta) \mathds{1}_{\Gamma_5 \otimes \Gamma_6} \\
    \epsilon &= E_{4f^2} - E_{4f^3}\\
    \mathbf{\Delta}^c(i \omega_n, \vk) &= \mathbf{V}(\vk)^\dagger \frac{1}{i \omega_n - \mathbf{\epsilon^{\text{imp}}}  - \mathbf{\Sigma}(i \omega_n) } \mathbf{V}^\dphant(\vk)\\
    \mathbf{V}^\dphant(\vk) &= \left( V_{0} \mathds{1}_{\Gamma_3 \otimes \Gamma_6} , V_{1} \mathbf{\Xi} \right)^T \mathbf{\ctrans}(\vk)\\
    \mathbf{\Delta}^{\text{imp}}(i \omega_n, \vk) &= \sum_{\vk} \mathbf{V}^\dphant(\vk) \mathbf{G}^{0}_{\text{eff}}(i\omega_n, \vk) \mathbf{V}^\dagger(\vk),
\end{align}
where bold quantities are to be understood as matrices in the respective conduction-electron or impurity quantum numbers. Note that the hybridization matrix $\mathbf{V}(\vk)$ is a $10 \times 2$ matrix at each $\vk$-point, mapping from effective conduction-electron space (i.e.~projected onto a specific set of \ce{V} d-orbitals) to the impurity space.

Due to the approximations mentioned in the main text (namely, the use of the effective conduction-electron propagator rather than the orbital-resolved one), it was not possible to implement a charge self-consistency loop between DFT and  DMFT-NCA.
%
%
\section{Magnetic Susceptibility}
The Pr 4f $\ket{\Gamma_6}$ and $\ket{\Gamma_5}$ states, as well as the conduction (Bloch) electrons, couple linearly to an external magnetic field $B$, while the $\ket{\Gamma_3}$ states couple quadratically in $B$. This is expressed in our Hamiltonian, keeping only linear terms in $B$ via
\begin{align}\nonumber
    H_{\mathrm{mag}} &= g_c B \sum_{n, \sigma, \vk} \mathrm{sgn}(\sigma) \, c^\dagger_{n,\sigma}(\vk) c^\dphant_{n,\sigma}(\vk)\\ \nonumber
    & \phantom{=} + g_d B \sum_{j, m} \mathrm{sgn}(m) \, d^\dagger_{j,m} d^\dphant_{j,m}\\
    & \phantom{=} + g_b B \sum_{j, \ichan} \mathrm{sgn}(\ichan) \, a^\dagger_{j,\ichan} a^\dphant_{j,\ichan}, 
\end{align}
with the g-factors $g_c = 2$, $g_d = 2$, and $g_b = -5/7$, calculated via $g_i = g_{J,L,S} \braket{J_z}_{i,+}$, where $g_{J,L,S}$ is the Land\'e-factor and the expectation value is taken with the state with a positive quantum number, i.e. $+1$ and $\uparrow$. Conduction electrons are coupled on the level of individual bands, where $g_c=2$ is assumed. The effect of the conduction electron Zeeman splitting on the impurity is weak compared to the direct coupling.

Although the $\Gamma_3$ response is enhanced by the Kondo effect, it will only give a sizable contribution once the field-induced splitting is comparable to the temperature $T$. 
The linear magnetic susceptibility is calculated (and measured) in the limit of small fields, for which we can neglect this contribution due to the $\sim B^2$ scaling of the Zeeman term. The impurity response is then given by direct contributions from $\Gamma_6$ and $\Gamma_5$. Noteworthy, the response of $\Gamma_6$ is a magnetic-spin, i.e., channel susceptibility. The conduction electrons produce a constant Pauli susceptibility.

\begin{figure}[tb]
    \centering
    \includegraphics[width=\columnwidth]{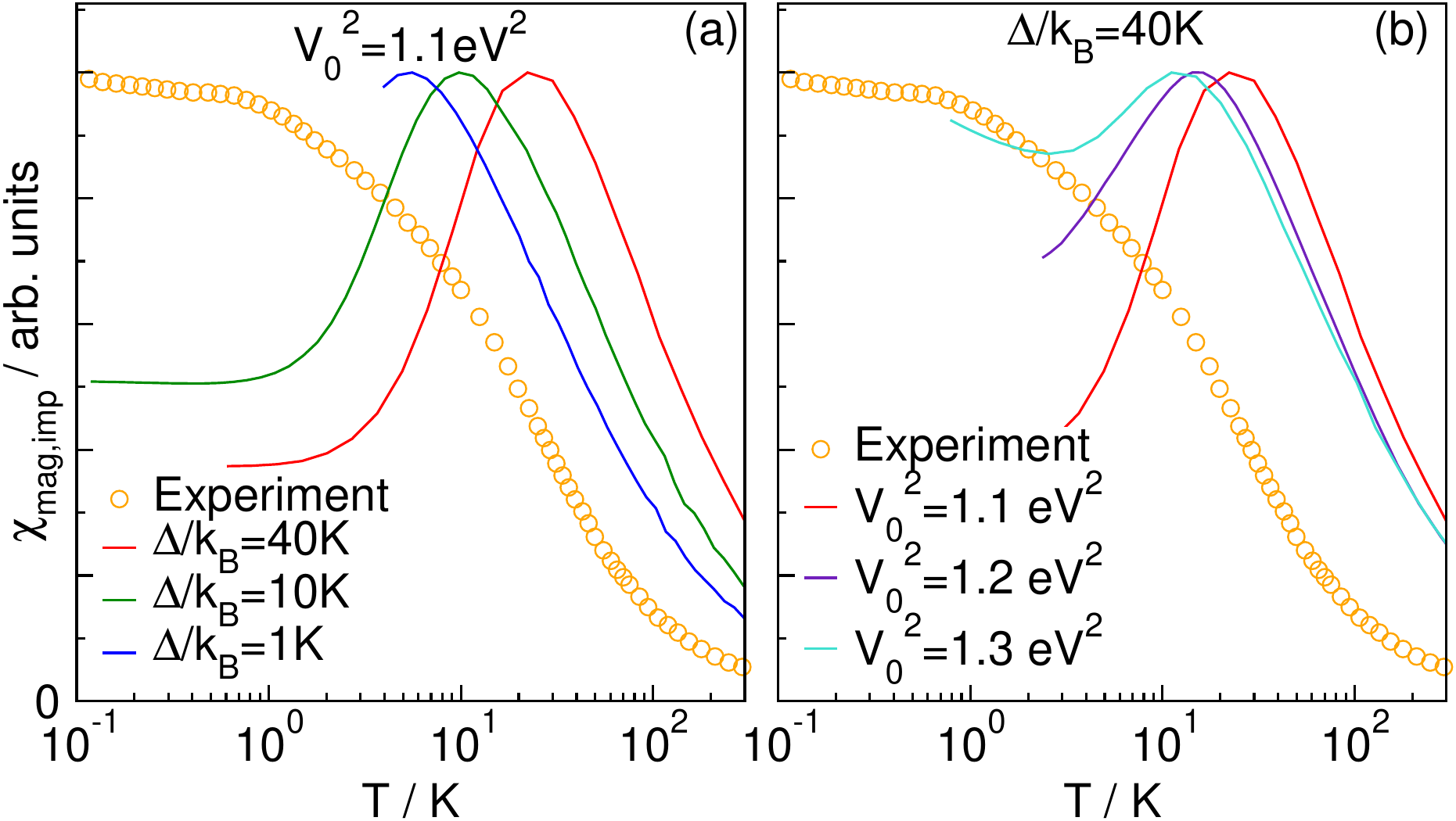}
    \caption{
        Impurity susceptibility for different microscopic parameters. In panel a), the hybridization is kept constant at $V_0^2 = \SI{1.1}{\eV^2}$, while the microscopic CEF splitting is modified. The microscopic parameter is, in general, not equal to the final, temperature-dependent CEF splitting. Panel b) shows the data for constant microscopic CEF splitting $\Delta/k_B = \SI{40}{\kelvin}$ but varied hybridization strength. Both panels show experimental data extracted from \cite{supp_araki2014}.}
    \label{fig:suscept_comparison}
\end{figure}

As discussed in the main text, the linear magnetic susceptibility was calculated by coupling the dipole moments of all involved states to a small magnetic field $B$ and taking the numerical derivative $\chi(T)=\partial M(T,B)/\partial B$ of the resulting magnetization $M(T,B)$. This includes all vertex corrections that would appear in a diagrammatic evaluation of $\chi(T)$ and that would arise from higher-order couplings between the Pr states. 

In Fig.~(4) of the main text, a qualitative agreement between regimes appearing in theory results and experimental data can be seen. Optimizing the similarity is a matter of fitting parameters of the microscopic model. 
A comparison of different parameter sets is shown in Fig.~\ref{fig:suscept_comparison}, in which we set $V_1 = V_0 \sqrt{3/2}$. This choice allows a more direct link between the microscopical splitting $\Delta$ and the effective splitting $\Delta_{\mathrm{eff}}$ and affected features in response functions. As displayed in panel a) of Fig.~\ref{fig:suscept_comparison}, reducing the microscopic CEF-splitting moves the local maximum in the susceptibility to smaller temperatures. The dependency is, however, not linear, which can be understood by level repulsion. The closer the CEF states come, the more they will be repelled by each other by hybridization. A comparison between different hybridization strengths for constant microscopic splitting parameter can be seen in panel (b). A larger hybridization also leads to the local maximum moving to lower temperatures. This can not be done indefinitely, though, due to the aforementioned level repulsion becoming stronger for increased hybridization. Another effect of an increased hybridization is the increased Kondo temperature and an increased slope of the $-\log(T)$ 2CK part.
%
%
\section{Dependence of the results on \texorpdfstring{$V_1/V_0$}{V1/V0}}
The exact ratio of $V_1/V_0$ (c.f. Eqs.~(5) and (8) in the main text) is, in principle, computable by constructing the full local many-body wavefunctions and calculating their overlap. The angular part of this is known from Group theory and included in our theory. The radial part, however, would give the relative hybridization strengths and was not computed.

First of all, an overall prefactor was left as a parameter to allow for a compensation of some of the approximations made. This allows one to restore the correct energy scales in the simplified model. Then, the ratio of $V_1$ to $V_0$ should be left as a parameter due to the same reasons - we are neglecting the $\Gamma_7$ conduction electrons and should allow for changes in $V_1$ to compensate for this approximation. The free parameters are therefore:
\begin{enumerate}
    \item the microscopic crystal field splitting $\Delta$;
    \item the overall prefactor $V_0$;
    \item the ratio $V_1/V_0$.
\end{enumerate}
Due to simplicity, we chose $V_1/V_0 = 1$ and adjusted $V_0$ and $\Delta$ to reproduce the correct renormalized CEF splitting and Kondo temperature. Although the results are quantitatively dependent on all parameters, the qualitative results did not change for a series of tested parameters.

\end{document}